\documentclass[
reprint,
superscriptaddress,
 amsmath,amssymb,
 aps,
]{revtex4-2}

\usepackage{graphicx}
\usepackage{dcolumn}
\usepackage{bm}

\usepackage{siunitx}

\begin{document}

\preprint{APS/123-QED}

\title{Multi-kT/m Focusing Gradient in a Linear Active Plasma Lens}

\author{K.N.\ Sjobak}\email{k.n.sjobak@fys.uio.no}
\affiliation{Department of Physics, University of Oslo, 0316 Oslo, Norway}
\affiliation{CERN, CH-1211 Geneva 23, Switzerland}
\author{E.\ Adli}
\affiliation{Department of Physics, University of Oslo, 0316 Oslo, Norway}

\author{R.\ Corsini}
\author{W.\ Farabolini}
\affiliation{CERN, CH-1211 Geneva 23, Switzerland}

\author{G.\ Boyle}
\author{C.A.\ Lindstrøm}
\author{M.\ Meisel}
\author{J.\ Osterhoff}
\author{J.-H.\ Röckemann}
\author{L.\ Schaper}
\affiliation{Deutsches Elektronen-Synchrotron DESY, Notkestraße 85, 22607 Hamburg, Germany}

\author{A.E.\ Dyson}
\affiliation{Department of Physics, Clarendon Laboratory, University of Oxford, Parks Road, Oxford OX1 3PU, United Kingdom}

\date{\today}

\begin{abstract}
Active plasma lenses are compact devices developed as a promising beam-focusing alternative for charged particle beams, capable of short focal lengths for high-energy beams.
We have previously shown that linear magnetic fields with gradients of around 0.3~kT/m can be achieved in argon-filled plasma lenses that preserve beam emittance~[C.A.\ Lindstrøm et al., Phys.\ Rev.\ Lett. 121, 194801 (2018)].
Here we show that with argon in a 500~\si{\micro m} diameter capillary, the fields are still linear with a focusing gradient of 3.6~kT/m, which is an order of magnitude higher than the gradients of quadrupole magnets.
The current pulses that generate the magnetic field are provided by compact Marx banks, and are highly repeatable.
These results establish active plasma lenses as an ideal device for pulsed particle beam applications requiring very high focusing gradients that are uniform throughout the lens aperture.
\end{abstract}

\maketitle

Plasma-based technology promises to pave the way for more compact particle accelerators~\cite{joshi_plasma_2003}.
While much research has been directed towards the use of plasma technology for compact acceleration~\cite{cros_towards_2017,schulte_application_2016,lindstrom_staging_2016}, there has been less effort directed at compactifying the focusing elements of the accelerator, which is also crucial for achieving compact machines.
The development of plasma lenses, as discussed here, may establish a superior beam-focusing alternative for charged particle beams in high energy physics applications, as well as for other demanding pulsed-beam applications such as medical applications or photon sources~\cite{joshi_perspectives_2020,couprie_towards_2018,maier_brilliant_2020}.

Quadrupole magnetic lenses are the conventional choice for focusing of high-energy charged beams~\cite[Section 7.2]{alexander_wu_chao_handbook_2013}.
Permanent quadrupole magnets can be made compact and precise~\cite{becker_characterization_2009}, but the focusing gradient is currently limited to several 100~T/m~\cite{becker_characterization_2009,modena_design_2012}, and they have a limited tuning range.
Furthermore, a quadrupole defocuses the beam in the plane orthogonal to the focusing plane, requiring a lattice of several magnets to achieve an overall focusing effect.
Active plasma lenses~\cite{van_tilborg_active_2015,panofsky_focusing_1950} (APLs) generate a focusing magnetic field by passing a strong current through a plasma that is parallel to and overlapping with the beam axis, creating an axisymmetric focusing force.
Axisymmetric focusing is particularly important for capturing and re-focusing beams with high divergence and energy spread, e.g.\ a beam produced in a plasma wakefield accelerator~\cite{lindstrom_staging_2016,migliorati_intrinsic_2013}.
High gradient is important to minimize the length of the focusing device; e.g.\ for a 100~GeV electron beam, a 10~cm long plasma lens with gradient of 3~kT/m would have a focal length of 1~m, whereas the focal length for a quadrupole magnet of the same physical length would be $>10$~m.
Implementations of APLs using discharge capillary technology~\cite{spence_investigation_2000} have recently been developed~\cite{van_tilborg_active_2015,lindstrom_overview_2018}, and first uses as optical elements for accelerator research applications are reported~\cite{pompili_focusing_2018,steinke_multistage_2016,barber_compact_2020}.

A challenge for APLs are different sources of nonlinearity in the fields, mainly due to nonuniform plasma heating~\cite{van_tilborg_nonuniform_2017} and wakefields~\cite{lindstrom_analytic_2018}.
Recent work has shown that the these nonlinearities can be avoided, and that active plasma lensing with linear focusing forces can be achieved~\cite{lindstrom_emittance_2018,pompili_focusing_2018}, which is necessary to preserve beam emittance.
Historically, very high gradients have been reached in $z$-pinch-discharge APLs~\cite{autin_z-pinch_1987,christiansen_studies_1984}, however these devices did not produce a linear focusing magnetic field.
More recently gradients of $\approx3.5~\mathrm{kT/m}$ have been demonstrated without pinching, however this was in a nonlinear hydrogen lens with a small diameter of 250~\si{\micro m}~\cite{van_tilborg_active_2015}.

When increasing the magnetic field gradient in an APL, several technical and physical challenges must be overcome in order to produce a useful device.
Our experiment pushes the demonstrated gradients by developing compact  high voltage sources with fast and reproducible pulses, while still avoiding both the thermal nonlinearities seen for light gases such as helium ~\cite{bobrova_simulations_2001,broks_nonlocal-thermal-equilibrium_2005,van_tilborg_nonuniform_2017,rockemann_direct_2018}, and the pinching of the field-producing current stream in the plasma~\cite{bennett_magnetically_1934}.
Our results demonstrate that APLs may repeatably operate at multi-kT/m gradients, outperforming quadrupole magnets, and at the same time provide linear focusing forces.
The results therefore represents a key step for establishing the APL as a versatile and useful device for compact, pulsed-beam applications.

\section{Experimental setup}
\begin{figure}
    \centering
    \includegraphics[width=\linewidth]{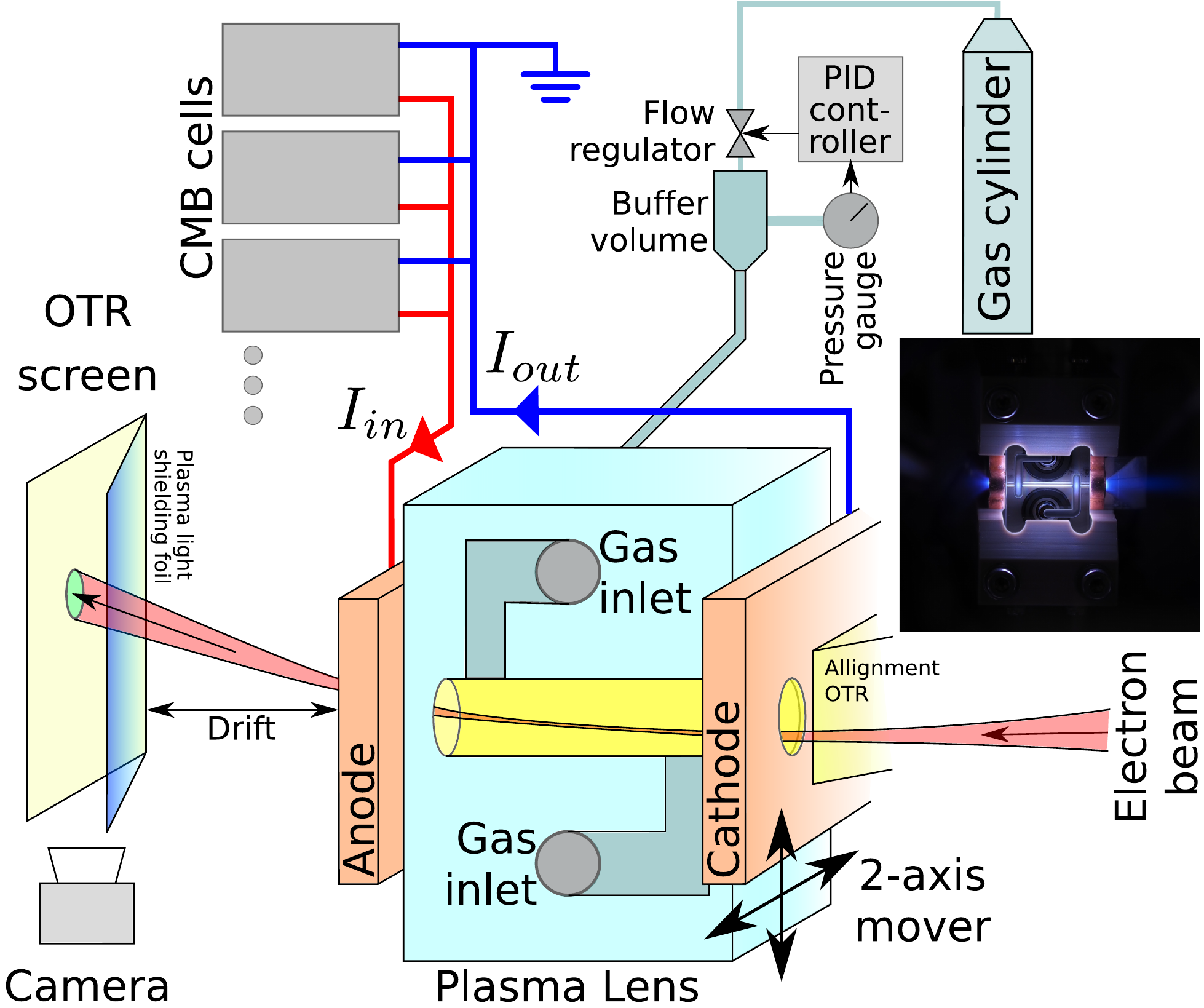}
    \caption{Overview of the most relevant components of the CLEAR Plasma Lens Experiment and the path of the beam.
             Insert: Plasma lens capillary during a discharge.
            }
    \label{fig:overview}
\end{figure}

The CLEAR Plasma Lens Experiment~\cite{lindstrom_overview_2018,lindstrom_emittance_2018,lindstrom_erratum:_2019} is installed at the CLEAR user facility at CERN~\cite{gamba_clear_2017, corsini_first_2018, sjobak_status_2019}, using a 200~MeV linear electron accelerator to produce beams with a large range of parameters.

The principle of the experiment is illustrated in Fig.~\ref{fig:overview}.
First, a strongly focused beam enters the capillary, in order to probe the field in a small region.
This beam is then deflected towards the center of the lens by the focusing force.
After exiting the lens, there is a short drift before the beam impinges on an optical transition radiation (OTR) screen mounted at a $45^\circ$ angle to the beam, used to measure the beam position.
This screen is shielded from the light emitted from the plasma with a thin metallized polymer foil; this limits us to only measuring vertical beam deflection, as otherwise the distance between the foil and the screen changes as the beam moves across it, causing a change in the apparent beam shape.

A small OTR screen is mounted on the upstream electrode in order to aid focusing the beam and aligning the lens to the beam, as shown in Fig.~\ref{fig:overview}.
The electron beam was focused down to a horizontal $\times$ vertical size of $44 \times 52$~\si{\micro m} RMS as measured on the alignment OTR and by the fall-off of beam intensity as a function of lens position when scanning the aperture edge over the beam.
This allows the beam to pass  cleanly through the capillary containing the current-carrying plasma and to sample the field locally.

In order to measure the magnetic field in the lens as a function of position, the lens is moved transversely relatively to the beam.
The beam position on the OTR screen is then measured as a function of the lens position, and compared to the beam position without discharge in order to measure the deflection.
The magnetic field gradient can then be reconstructed as discussed in Appendix~\ref{sec:elliptical}; due to the very high gradients, the plasma lens must be treated as a thick lens because the focal length ($\approx$ 20~mm from lens entrance) is on the order of the length of the lens.

Furthermore, the gradient enhancement was measured as a function of time, by holding the lens at a known position and changing the relative timing of the lens with respect to the beam.
The gradient enhancement was then found by comparing the deflection expected from the current measurement with the one observed on the screen, as described in detail in Appendix~\ref{sec:timeThick}.

\subsection{Discharge capillary}

The discharges are contained within a sapphire capillary, as shown in the insert in Fig.~\ref{fig:overview}.
For the experiments reported in this paper, a 15~mm long capillary with a nominal diameter of 500~\si{\micro m} was used.
It is made by milling out a semicircular groove in two sapphire plates which are then pressed together.
For small-diameter capillaries, the manufactured capillary may not be perfectly round; this is taken into account when computing the expected gradient from the current as explained in Appendix~\ref{sec:elliptical} as well as for modeling the nonlinearity in helium as described in Appendix~\ref{sec:JT-elliptical}.

The capillary is supplied with gas at an adjustable pressure through two inlet channels, giving a flat initial pressure profile between these channels.
In order to maintain the accelerator vacuum, a turbo-pump removes the gas escaping from the capillary openings, and a 3~\si{\micro m} Mylar beam window is used to separate the plasma lens chamber from the upstream accelerator vacuum.

The breakdown voltage and drive current pulse is supplied via a set of copper electrodes on each end of the capillary, with holes that allow the beam to pass in and out of the channel.

\subsection{Compact Marx Bank}

\begin{figure}
    \centering
    \includegraphics[width=\linewidth]{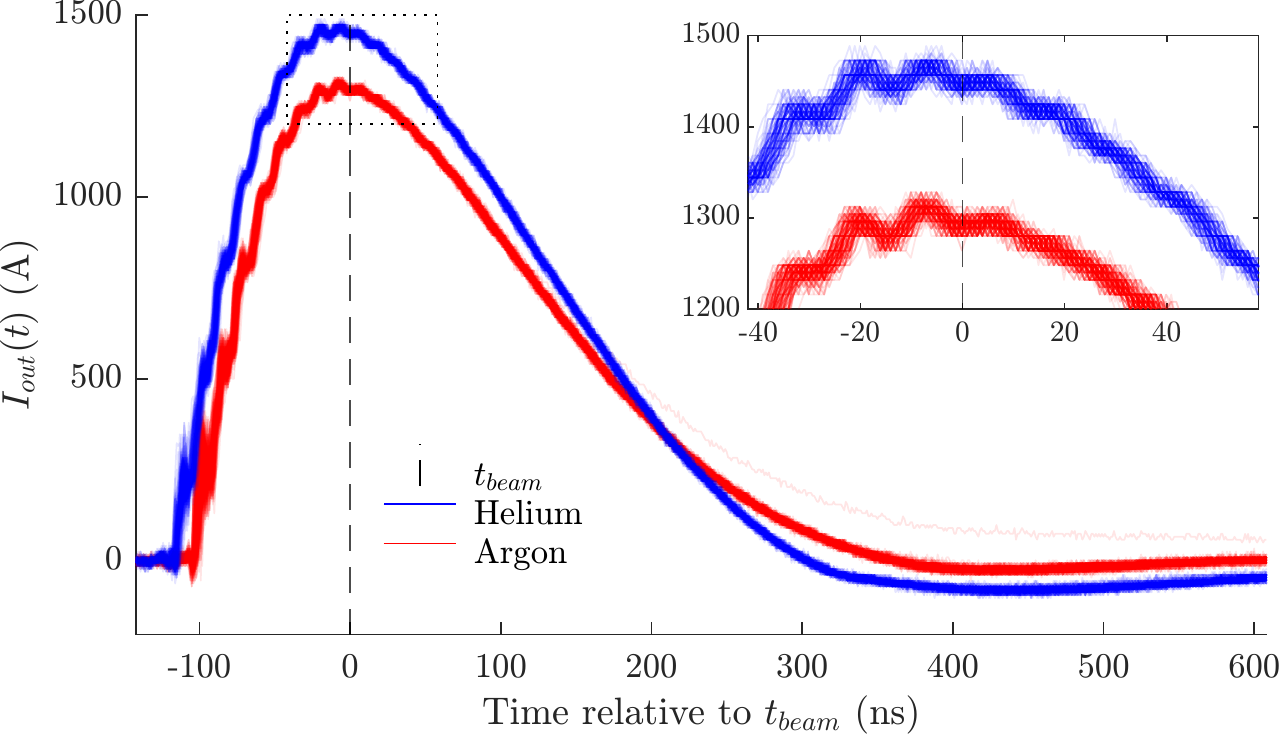}
    \caption{Compact Marx bank current traces for the offset scans in helium and argon that are shown in Figs.~\ref{fig:gradient-Ar} and \ref{fig:gradient-He}.}
    \label{fig:CMBcurrents}
\end{figure}

The high voltage required to first form the conducting plasma and then drive the high currents needed for our high-gradient lens were provided by Compact Marx Bank (CMB) cells~\cite{dyson_compact_2016}.
These were connected in parallel, with each CMB cell consisting of a 10-stage ladder network with $2\times22~\mathrm{nF}$ capacitor rungs initially charged to $2.0~\mathrm{kV}$.
This gives a floating initial discharge voltage of $20~\mathrm{kV}$ with an effective cell capacitance of 4.4~nF (about 1~J per discharge).
This voltage is higher than that strictly needed for gas breakdown, since this ensures a rapid onset of the discharge within a few ns of the voltage pulse being applied.
Once the plasma is formed the current pulse starts to climb, mainly constrained by circuit inductance.

The CMB cells are arranged on a single chassis in banks of four and enclosed in a $210 \times 297 \times 50~\mathrm{mm}$ box.
Up to 8 cells were used, placed a within a few~\si{\centi m} from plasma lens.
The cells provide high-current pulses with sub-\si{\micro s} duration and fast rise time as shown in Fig.~\ref{fig:CMBcurrents}, appropriate for driving a plasma lens while limiting the total amount of energy deposited in each pulse to reduce capillary damage by heating.
The close proximity of the mounting is necessary to eliminate the need for impedance matching.

Combining several cells in parallel brings a number of advantages.
Firstly the total current is increased according to the number of cells used and thus the gradient.
Simultaneously the jitter in both timing and peak current is reduced.
Furthermore, the reliability is also increased, as the current drawn from each cell is reduced to below 200~A, and in the case of the loss of a single cell the lens can still operate in a degraded peak current mode.

The current output of the CMBs are measured by two current-pulse transformers; one on each cable in order to detect large losses to ground.
The current traces during the offset scans described in the next section are shown in Fig.~\ref{fig:CMBcurrents}.
This shows that the current rises from 0 to peak in $\approx100$~ns, has a flat-top of a few 10s of ns, before it falls off in $\approx300$~ns.
For both argon and helium the incoming and outgoing currents at the time of beam arrival are identical, and their standard deviation are both equal to the sampling resolution, which is 8~A.
In general the current pulses are highly reproducible.

\section{Magnetic field gradient measurements}

\begin{figure}
	\centering
	\includegraphics[width=\linewidth]{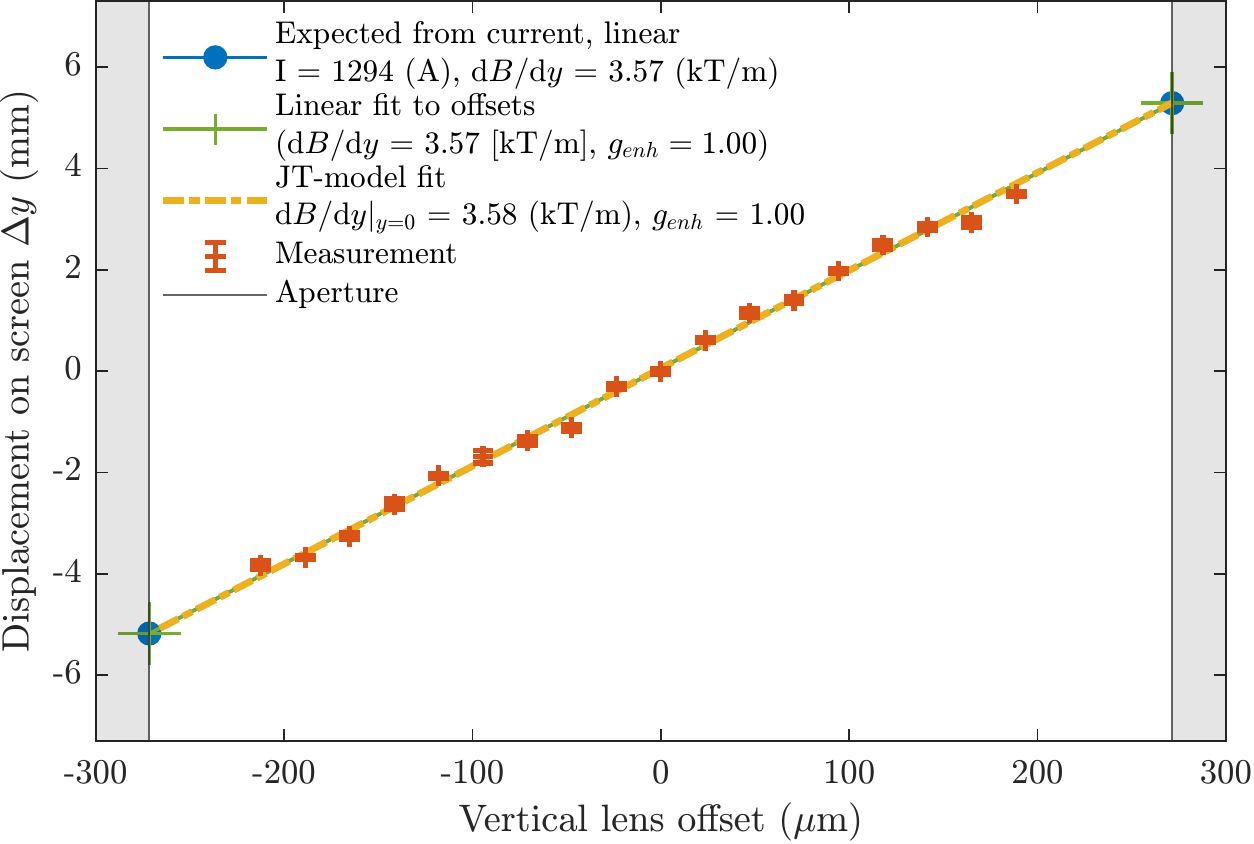}
	\caption{Beam displacement as a function of lens offset in argon on peak current, for measuring the field gradient.}
	\label{fig:gradient-Ar}
\end{figure}
\begin{figure}
	\centering
	\includegraphics[width=\linewidth]{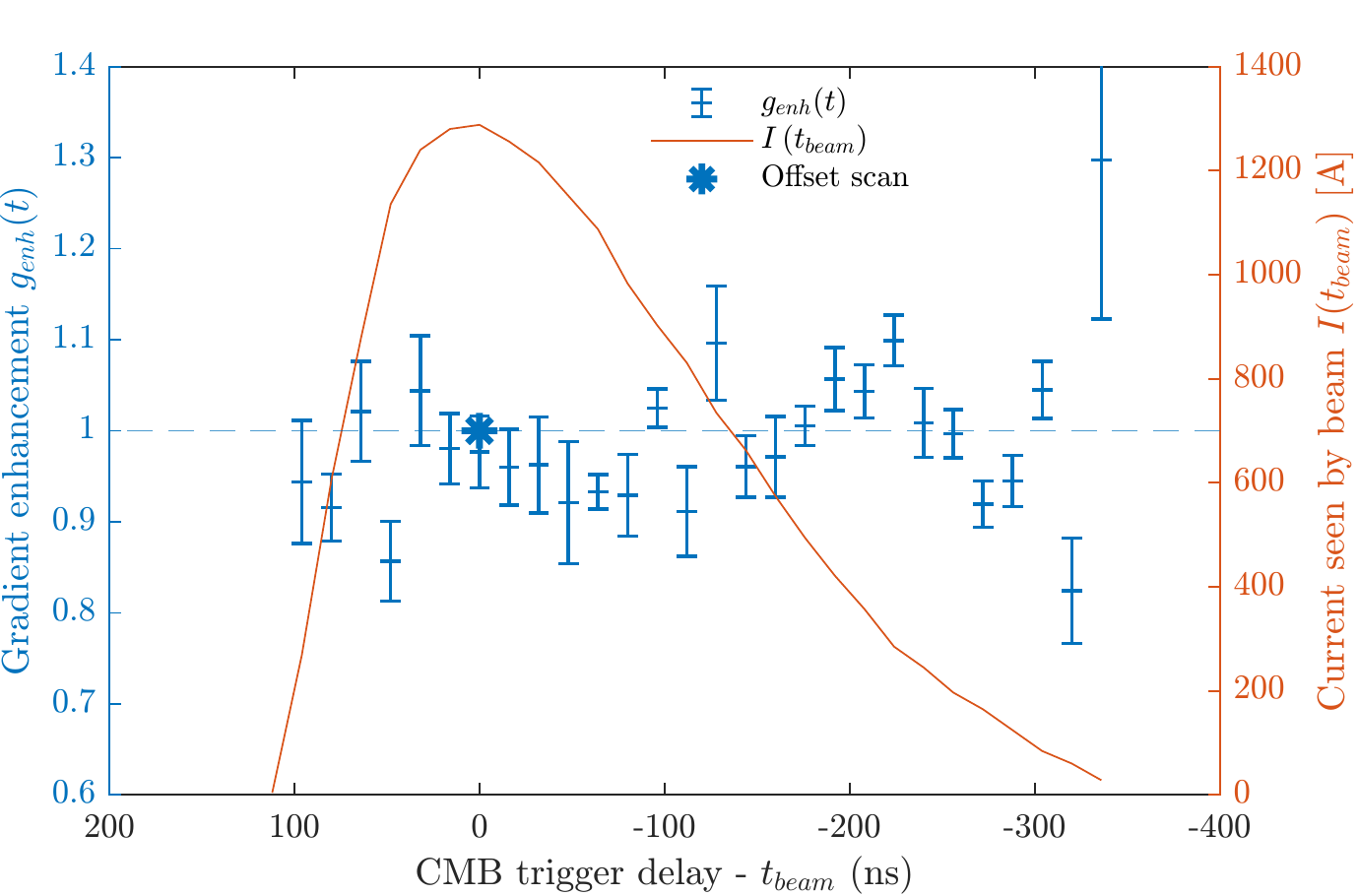}
	\caption{Estimated gradient enhancement as a function of time at a fixed offset ($y_0=-150~\si{\micro m}$) in argon.
	         The result from the offset scan shown in Fig.~\ref{fig:gradient-Ar} is also indicated.}
	\label{fig:time-Ar}
\end{figure}
\begin{figure}
	\centering
	\includegraphics[width=\linewidth]{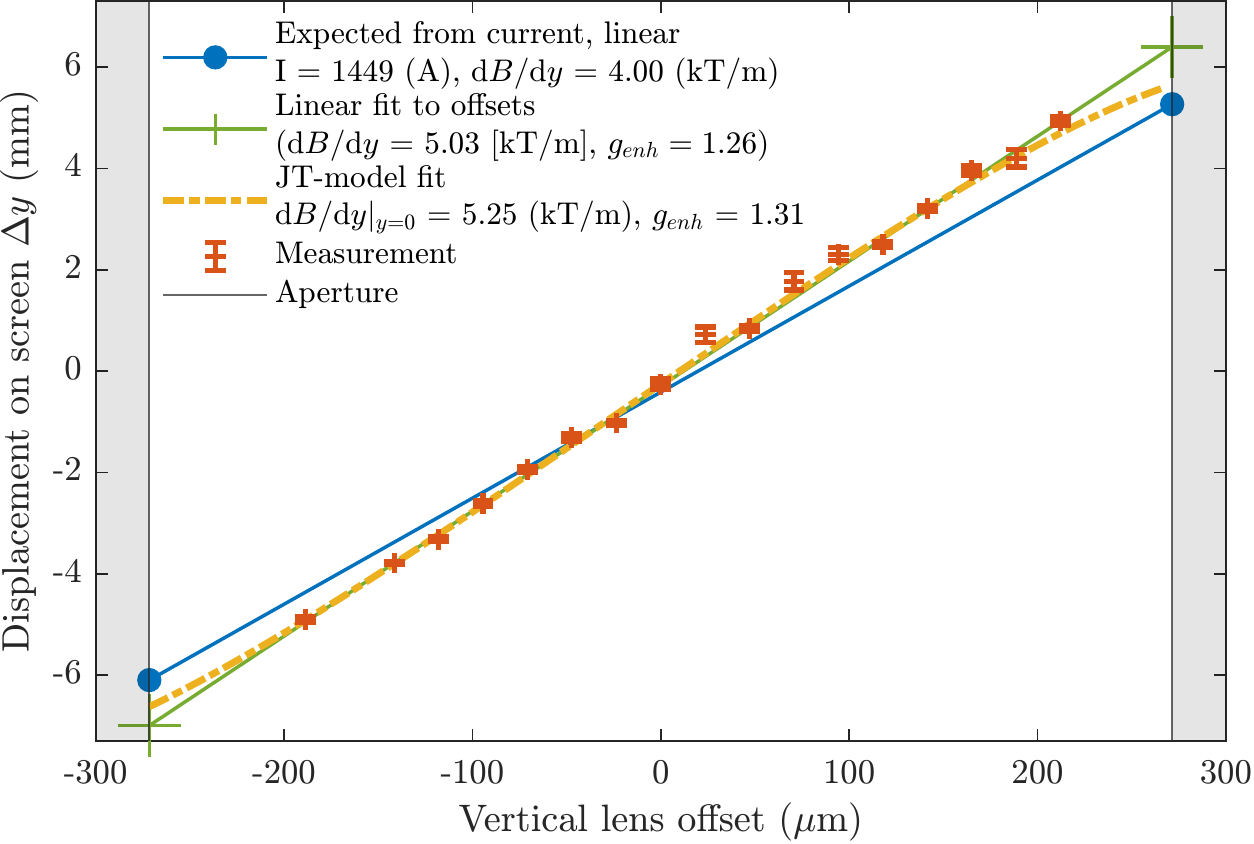}
	\caption{Beam displacement as a function of lens offset in helium on peak current, for measuring the field gradient.}
	\label{fig:gradient-He}
\end{figure}

The field measurements were done in both argon and helium, and the lens was operated at a neutral gas pressure in the buffer volume of 25~mbar and 55~mbar respectively.
These were the minimum values for stable operation of the discharge in the 0.5~mm capillary used here.
The expected pressure in the capillary is then approximately 8~mbar for argon and 17~mbar for helium~\cite{lindstrom_emittance_2018}.
The beam energy was 195~MeV, using three bunches for a total train length of 2~\si{\nano s}, each with a bunch length 4~ps~rms, and a total charge of 200~pC.
The beam size on the screen soon after the current had returned to zero was the same as before plasma ignition, indicating negligible influence from plasma wakefields~\cite{lindstrom_analytic_2018}.
This relatively high beam charge was necessary to be able to measure the beam position on the OTR screen in spite of the strong over-focusing in the lens, causing a diffuse beam with a size of $\approx1.4~\mathrm{mm}$ at this location.
The large size made the accurate reconstruction of the beam position more difficult, contributing to the errorbars seen in Figs.~\ref{fig:gradient-Ar} and~\ref{fig:gradient-He}, which indicate the standard error of the mean reconstructed beam position.
The field was measured only in the central 400~\si{\micro m} of the capillary; in the last $\approx 50~\si{\micro m}$ from the edge the beam was deflected outside the screen.

The beam displacement as function of position for argon is shown in Fig.~\ref{fig:gradient-Ar}.
The measured gradient from a linear fit was found to be $3.57\pm0.08$~kT/m, where the error estimate is the standard deviation due to the uncertainty of the position measurements.
No evidence of a gradient enhancement~\cite{van_tilborg_nonuniform_2017} near the center was observed when fitting a nonlinear model.

The measurement of gradient enhancement as a function of time is shown in Fig.~\ref{fig:time-Ar}.
This shows that the gradient enhancement remained constant throughout the discharge; the weighted average being $0.99\pm0.08$.

For helium the beam displacement as function of position is shown in Fig.~\ref{fig:gradient-He}, where the observed central gradient was $5.25\pm0.13$~kT/m.
Here the nonlinear model found a gradient enhancement of $31\pm3$\%, where the error is the standard deviation due to the uncertainity of the position measurements.

\section{Discussion}

The central focusing gradients were obtained directly from the beam deflection measurement, using fits based on the measured current and geometry and the corrected mover positions.
For argon the results indicate that even at these very high gradients, APLs remain linear throughout the time- and aperture range that we tested, which is a necessary condition for emittance preservation.
For helium the observed gradient enhancement factor is statistically consistent with the 34\% found earlier~\cite{lindstrom_emittance_2018,lindstrom_erratum:_2019} at a central lower gradient 440~T/m, using a 1000~\si{\micro m} diameter capillary and single CMB cell producing a peak current of 410~A.

The gradient enhancement factor in argon was also measured as a function of time, confirming that it remains constant and close to unity throughout the duration of the discharge.
This implies that the gradient can be fine-tuned simply by changing the timing of the discharge, without incurring nonlinearity.

In earlier work~\cite{lindstrom_analytic_2018} it has been shown that beams with high charge density can create waves in the plasma, which in turn may lead to nonlinear fields.
The relative effect of the wakes may be reduced by increasing the plasma lens gradient~\cite{carl_andreas_lindstrom_emittance_2019}, hence our results increase the parameter range where APLs with linear fields can be used.

Another potential source of nonlinearities is the $z$-pinching of the current in the plasma lens, due to the magnetic pressure overcoming the gas pressure in the plasma.
This will tend to concentrate the current on axis, forming a nonlinear and potentially unstable field from the concentrated current~\cite{bennett_magnetically_1934,rosenbluth_infinite_1955}.
We did not observe any indication that this was the case here, possibly due to the short duration of the peak current.

The group of experimental CMB cells reached very high currents without any major failures.
The addition of several of these in parallel tended to stabilize the discharge compared to previous experiments, and reduced the failure rate of the system as less current was drawn from each CMB cell.
This principle allows for coarse tunability of the total current and redundancy in the system by overprovisioning the number of cells needed for the nominally required current.

While the inner surfaces of the capillaries appeared darker after carrying discharges with $> 1~\mathrm{kA}$ currents, we did not observe any degradation of performance during the experiment.
However, before application of high gradient plasma lenses more systematic studies of the capillary damage should be performed in order to ensure long-term stability.

\section{Conclusion}

We demonstrate the operation of a compact, linear APL in argon with a focusing gradient of 3.6~kT/m.
With helium as the fill gas the lens is nonlinear with a central gradient of 5.3~kT/m.
The lens has a 500~\si{\micro m} diameter and is powered by compact Marx banks (CMB) cells connected in parallel.

The use of CMB cells enable us to reach the high currents required to drive the lens with short-duration pulses.
This presents compact power source that can reach even higher currents by simply adding more cells.

This opens up the possibility for the use of APLs in applications requiring gradients well beyond those of quadrupole magnets, or that otherwise benefit from a strong axisymmetric linear focusing element.

\begin{acknowledgments}
We thank CERN for providing beam time at the CLEAR User Facility, and especially A.\ Gillardi, D.\ Gamba, A.\ Curcio, A.\ Chauchet, S.\ Curt, G.\ McMonagle, S.\ Burger, M.\ Bergamaschi, T.\ Lefevre, E.\ Granados, and H.\ Panuganti, for making it possible to run the accelerator, and for often lending us a hand and suggestions for improving the experiment.
We also thank S.\ Hooker for suggestions for improving the manuscript.
This work was supported by the Research Council of Norway (NFR Grant No.\ 255196/F50) and the Helmholtz Association of German Research Centers (Grants No.\ VH-VI-503 and No.\ ZT-0009).
\end{acknowledgments}

\newpage
\appendix

\section{Magnetic field in elliptical capillaries}
\label{sec:elliptical}

The capillaries are produced by milling sapphire plates using a ball mill cutter with the same diameter as the capillary, which should plunge exactly 1/2 of the milling bit diameter.
In practice the machining precision can be at a level where small diameter capillaries are noticeably oval.
This distributes the current over a larger area, causing a reduction of the gradient that is different between the horizontal- and vertical planes.
For the capillary used in this paper, the diameters were measured using optical profilometry to be 525~\si{\micro m} horizontally and 543~\si{\micro m} vertically.

Working in an elliptical coordinate system $x=\xi a\cos\theta$, $y = \xi b\sin\theta$ where $a,b$ are the elliptical half-axis and $\xi \in [0,1]$, we will make the assumption that both the magnetic field component $B_\theta$ and the current density $J$ is independent of the polar angle $\theta$.
Applying Ampere's law on loops of constant $\xi$, we find that
\begin{equation}
    B_\theta(\xi) = \frac{\mu_0 2\pi a b}{P(\xi)} \int_0^\xi J(\xi') \xi'~\mathrm{d}\xi'\,,
    \label{eq:ampere}
\end{equation}
where $P(\xi)$ is the circumference of an ellipse with the same flattening $\varepsilon = (a-b)/a$ as the capillary itself, which can be estimated with Ramanujan's formula
\[
P \approx \pi \xi \left[ 3 (a+b)-\sqrt{(3a+b)(a+3b)} \right]~.
\]

For the case of a uniform current density, the expected horizontal field gradient in an elliptical capillary is then
\begin{equation}
	g_h = \frac{\mathrm{d}B_y}{\mathrm{d}x} \approx \frac{B_\mathrm{surf}}{a} \approx \frac{\mu_0 I}{a P(\xi=1)}\,
	\label{eq:ampere-gradientUniform}
\end{equation}
and similar for the vertical plane by substituting $a \to b$, where $B_\mathrm{surf} \equiv B_\theta(\xi = 1)$.
This approximation was used when computing the expected gradient $g_\mathrm{uniform}(I;a,b) = g_h \;\mathrm{or}\; g_v$.
Through simulation, it was verified that for a flattening $\varepsilon = (a-b)/a \le 0.5$ and a uniform $J$, the agreement between the resulting analytical formula in Eq.~\ref{eq:ampere-gradientUniform} and the simulation was better than 1\%.

\section{$J \propto T^{3/2}$ model in elliptical capillary}
\label{sec:JT-elliptical}

The JT-model~\cite{van_tilborg_nonuniform_2017,bobrova_simulations_2001,lindstrom_emittance_2018} describes how a gradient nonuniformity can be caused by a radial variation of the electron temperature $T_e = A u(\xi)^{2/7}$ in the capillary, where $A$ is a constant given by the capillary size and plasma parameters, and $u$ is a solution to the heat flow equation
\begin{equation}
    \frac{1}{\xi} \frac{\mathrm{d}}{\mathrm{d}\xi} \left( \xi \frac{\mathrm{d}u}{\mathrm{d}\xi} \right) = - u^{3/7}\,.
    \label{eq:diffU}
\end{equation}
This causes a variation of the conductivity $\sigma = \sigma_0 T_e^{3/2}$, which causes the nonuniformity of the current density $J=\sigma E$, creating the nonlinearity of $B_\theta$.
Combining these gives $J = \sigma_0 E A^{3/2} u^{3/7}$, where the unknown constant $\sigma_0 E A^{3/2}$ can be determined through conservation of current $I = \int_0^1 J(\xi)\,\mathrm{d}\xi$, and $u(\xi)$ through solving Eq.~\ref{eq:diffU}.

If the flatness $\varepsilon$ of the capillary is small, it is reasonable to assume that $u$ and thus the current density is not a function of $\theta$.
The current density is then given as
\begin{equation}
    J(\xi) = \frac{I u(\xi)^{3/7}}{2\pi a b \int_0^1 u^{3/7} \xi\,\mathrm{d}\xi}
    \label{eq:JT}
\end{equation}
which for a round capillary where $a=b=R$ reduces to the expression in~\cite{van_tilborg_nonuniform_2017}.
The magnetic field can then be determined through Eq.~\ref{eq:ampere}.

\section{Magnetic field gradient measurement technique considering thick-lens effect}
\label{sec:thickLens}

\begin{figure}
	\centering
	\includegraphics{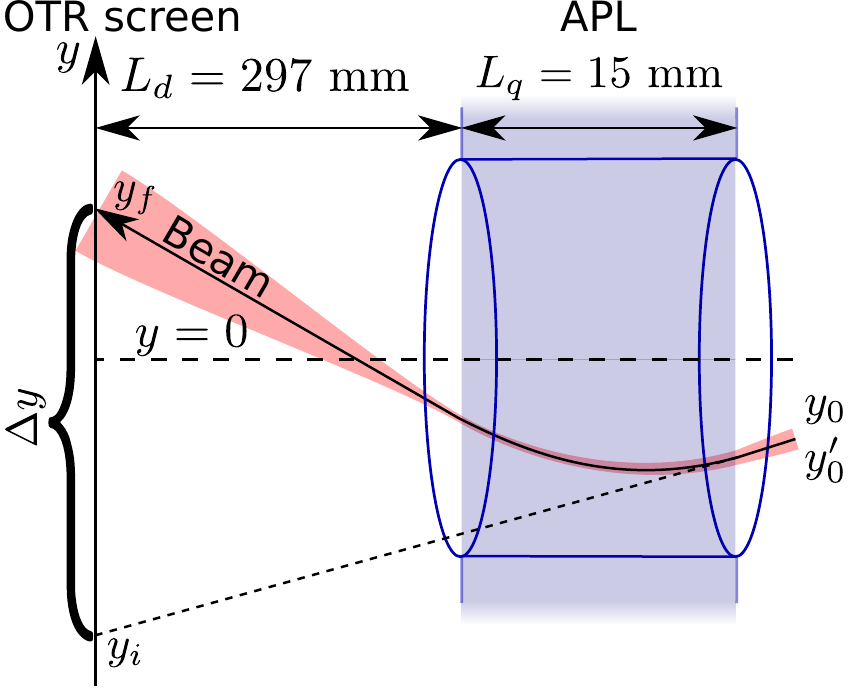}
	\caption{Definition of variables used in discussion of thick-lens gradient reconstruction.}
	\label{fig:measurement_principle}
\end{figure}

The measurement of the magnetic gradient is based on measuring the deflection of a particle beam passing through the lens, as illustrated in Fig.~\ref{fig:measurement_principle}.
The position of the beam on the downstream profile monitor is compared between the case where the plasma lens is on and off, giving a deflection $\Delta y = y_f-y_i$.

Because the focal length of the lens is short compared to the length of the lens, the thick-lens effect must be taken into account to correctly find the gradient.
The position on the screen is given by
\begin{equation}
\begin{split}
y_f &= y_0 \left(\cos(\sqrt{K}L_q) - \sqrt{K}L_d\sin(\sqrt{K}L_q)\right) \\
    &+ y'_0 \left( \sin(\sqrt{K}L_q)/\sqrt{K} + L_d \cos(\sqrt{K} L_q) \right) \\
    &\equiv y_0 A + y'_0 B\;,
\end{split}
\label{eq:yf}
\end{equation}
where the focusing strength is
\[
K \mathrm{[1/m^2]} = \frac{\mathrm{d}B_y}{\mathrm{d}x} \mathrm{[T/m]} \frac{c \mathrm{[m/s]}} {P_0 \mathrm{[MeV]} 10^6}~.
\]
The deflection on the screen is thus given as
\begin{equation}
\begin{split}
\Delta y &= y_0 A(K) + y_c - y_i(y_0) \\
	     &= y_f(K,y_0,y_0') - y_i(y_0,y_0')\;,
\end{split}
\end{equation}
where $y_c$ is a constant term due to the initial beam angle $y'_0$, and $y_i = y_0 + y_0'(L_q+L_d)$ is the position on the screen when $K \to 0$.
Note that in the experiment the lens itself was moved relative to the beam, which mathematically is equivalent to shifting the point where $y=0$ and thus shifting both $y_0$ and $y_f$ in the opposite direction of the lens movement.

In order to find the strength $K$ and thus the gradient seen by the beam $g_\mathrm{beam} = \frac{\mathrm{d}B_x}{\mathrm{d}y}$, we measure the deflection as a function of the initial position  $\mathrm{d}\Delta y/\mathrm{d}y_0$ and numerically find the root of
\begin{equation}
\frac{\mathrm{d} \Delta y}{\mathrm{d}y_0} - A(K) + 1 = 0\;.
\end{equation}

The method was tested by reversing the current direction in the lens, running it in the defocusing mode with argon gas.
In this case, the beam is pushed outwards while traversing the lens, increasing the kick instead of decreasing it; however the same focusing gradient was measured in the reversed case as in the normal case when other parameters were kept constant.

In order to apply this method to a nonlinear thick elliptical lens with a field distribution given by Eqs.~\ref{eq:JT} and~\ref{eq:ampere}, a numerical tracking method is used instead of equation~\ref{eq:yf} to track the beam centroid through the lens.
For this, the lens is sliced into 1~mm thick slices, and the field seen by the beam centroid is computed at each slice.
The computed deflection is then fitted to the data, using the initial angle $y'_0$ and the central normalized temperature $u(0)$ as free parameters, and the measured current and geometry as input.

\section{Field enhancement time development considering thick-lens effect}
\label{sec:timeThick}

For measuring how the gradient enhancement $g_\mathrm{enh} = \frac{g_\mathrm{beam}}{g_\mathrm{uniform}}$ develops in time, timing scans are used.
In these, the trigger of the CMBs are scanned relative to the time of arrival of the beam so that the beam probes the discharge at different times.
For these experiments, the position of the lens and the initial beam position is kept constant; if assuming a linear lens the centroid position on the final screen is then given by Equation~\ref{eq:yf}.
This depends on three parameters: The focusing strength $K$, and the initial beam position $y_0$ and angle $y'_0$ relative to the lens axis.
In the experiment $y_0$ is known as the lens mover is well calibrated when ran without discharge, and the center is found as the lens position where the centroid kick is zero, and verified as being in the center of the aperture.
This procedure works well as long as $y'_0$ is small, which is ensured during beam setup by varying the beam angle in order to maximize the apparent beam aperture through the capillary.

By measuring the beam deflection $\Delta y$ at several $K(I(t))$ and assuming that $g_\mathrm{enh}$ is constant and $y_0=-150~\si{\micro m}$, the initial beam angle and the overall gradient enhancement can be found through a nonlinear least-squares fit, finding $y_0' = 0.85~\mathrm{mrad}$ and $g_{enh} = 0.98$.
The angle is also small enough that the reduction of apparent aperture over the length of the 15~mm long capillary would be barely noticeable; it is also robust as if locking $g_{enh}=1$ and fitting $y_0$ and $y'_0$, a position of --147~\si{\micro m} and an angle of --0.84~mrad.

Once the initial position and angle is established, the actual strength $K$ and thus the $g_\mathrm{beam}$ can be found for each time-point by solving Equation~\ref{eq:yf} for $K$ and computing $g_\mathrm{beam}$.
Comparing this to the expected gradient in case of a uniform current then gives an estimate of the gradient enhancement $g_\mathrm{enh}$ as a function of time, as shown in Fig.~\ref{fig:time-Ar}; even if the original fit was made under the assumption of $g_\mathrm{enh} = 1$, a time development would still be expected to be apparent.
The errorbars on the gradient enhancement is the estimated standard deviation derived from a combination of 1~ns uncertainty on the time of sampling the current for the calculation of $g_\mathrm{uniform}$, and the uncertainty on the mean beam position on the screen at each timing for the calculation of $g_\mathrm{beam}$.

\end{document}